Short Paper*

# Impact of Critical and Auto Ticket: Analysis for Management and Workers Productivity in using a Ticketing System

Kent Darryl M. Aglibar
Graduate Studies, University of the East, Philippines

Nelson C. Rodelas
Graduate Studies, University of the East, Philippines
(corresponding author)

## Abstract

*Purpose* – Ticketing system is common in Technical Support in Information Technology Industry. At present time, even management is using it. It serves as a way to connect the company and the client, end to end. Support Industry used this also to monitor and log all solutions and processes that come up to resolve the ticket. Client and the resources that will troubleshoot this, use the ticket to have an idea for what is the problem, when the client encountered it, and even the business impact. The researchers conducted research entitled "Impact of Critical and Auto Ticket's: Analysis for Management and Workers Productivity in using Ticketing System" where it aims to come up with a solution on how we are going to prevent, troubleshoot, and give insight for possible business impact to those everyday issues. This will help the management's business process to have an analysis to prevent critical issues that will affect the live production as well as give time to their resource to focus on their tasks.

*Method* – Researchers used data collection to gather data from management, support workers, and Service Now's open-source ticketing system to visualize the ticketing system application. Also, a purposive sampling type of survey has been used to highly represent the data.

*Results* – Critical ticket gives a lot of pressure to the resources as they needed to resolve the incident in accordance with the service level agreement. It affects their productivity to maximize their time to resolve other escalated incidents as they need to focus with critical ones. Instead of maximizing time management and prioritizing employee's welfare like work and life balance, auto ticket's affects it because of continuous re-occurring and sometimes needed to be escalated as it will affect the live production. Having knowledge management helps resource to find references on how to deal with the incident. It helps them to execute workaround quickly and think of a way on how to resolve it permanently.

*Conclusion* – It is concluded that critical and auto ticket's affects the everyday productivity of the worker especially teaching new employees despite ongoing critical incidents. Researchers provided solutions such as knowledge Management and Dashboard to document all the solutions encountered and monitor the SLA's and incoming tickets.

*Recommendation* – It is recommended to have further research on how critical and auto ticket affects the mental health of resources and its direct impact to businesses. It is also recommended to have a study on how knowledge management work and help resources to identify correct workaround despite of having a lot of troubleshooting guides.

*Practical Implication* – This research benefits data warehousing companies that can be used as a third-party application to connect support team, management, and clients. Management can create a dashboard and identify how their support team provides good quality of service with their response and resolution time. Support team can have a reference to quickly fix escalated incidents and provide ticket number to the client for easily follow up an update.

## INTRODUCTION

Support Operation is very well known in Information Technology Field. Aside from having a web development team, network security, and any other IT Team, Support is one the most important role. Having a support team is also maintaining a good quality of the application and has a significant impact on business performance in terms of profitability (Stojanov, Dobrilovic, & Jevtic, 2011). One of the strategies of software companies in the present time is to get the attention of the market by using an application system that will provide the needs of the client. Because of this, it has a worldwide positive impact and countries using it improve themselves to a new IT infrastructure (Xinzhou, 2015). One of the benefits of the application system is it can be viewed on mobile which is clients or users can access it anywhere, everywhere, and anytime (Xinzhou, 2015). Warehousing companies nowadays are using application systems to manage remotely the day-to-day transactions. To supervise the entire business process, management is using Enterprise Resource Planning (ERP) Application as part of their IT strategy and model (Davis, 2011). The support team should provide day-to-day control for the application once it is ready for production (Bourne, 2014). Because it is a live production, errors may come unexpectedly that's why users always seek assistance to give a solution or ask something about their issues. With the said problem, the Support Team is using a Ticketing System to easily identify and recognize the possible issue.

There are two types of tickets, Incident, and Request Ticket. An incident ticket is a record of failures, errors, and troubleshooting that has been executed. It is also containing structured or non-structured data and what is the error encountered (Li et al., 2014). Request ticket is a record of users' requests to execute to the support team. With the help of the Ticketing System, it is easy for management to monitor the resolution time of the issue or it will serve as a reference if that issue comes again next time around. As per the research, as they did an experimental result on training, the information inside the ticket like work notes and description was one the main component that increase the model prediction accuracy from 53.8% to 81.4% (Al-Hawari & Barham, 2019).

Auto Ticket that has been received by the support team is normal to a 24 by 7, day-to-day transaction. It has been easy to troubleshoot because it frequently comes but unfortunately, not all days for the support team is stress-free day. When the users encountered a critical ticket, the support team troubleshoots it under the Service Level Agreement (SLA) and it is time-consuming (Bourne, 2014). Some Auto Tickets are no urgent business impact but if this has been too long, in some ways it creates a critical impact on the application users. Shipments, Receiving, Picking, Storage, and Transactions are examples of affected when it comes to critical tickets that will also contribute to a low percentage of the production's productivity. "They found out that all the operational issues are interrelated and mutually affect the performance of one another. De Koster et al. (2007)" (Al-Hawari & Barham, 2019).

Critical tickets also affect the databases and make the system slow that will contribute also to a low percentage of productions productivity (Khanzode & Shah, 2017). Low production productivity or production downtime causes losses to the company and loses the trust of the clients. The main purpose of the research is to provide an analysis on the impact of Critical and Auto tickets in our data warehouse, to provide some solution on how we are going to resolve auto tickets to avoid critical issues in following the service level agreement between the service provider and clients and to have an overview on some general problems that may cause a critical impact to the production.

**LITERATURE REVIEW**

*Service Now*

The Ticketing System is one of the popular tools in the IT Industry that connects both clients and companies. Using it, it is easy to monitor, check, and document all transactions that come every day. This also helps the business process as the user just needs to take note of the ticket number to follow up easily. It provides both ends the proper documentation and keeps track of the progress. Support teams are one of the main beneficiaries of this tool as they can use it to backtrack and as a reference, if the escalated report of the clients is re-occurrence (Gohil, & Kumar, 2019).

Service Now is one of the popular Ticketing Systems. It provides infinite data storage for all the transaction documentation. Aside from ticketing, it has also other features like storage of Knowledge Management, Dashboards, and Task boards that can help the team to find references in fixing escalated reports and monitoring the open ticket count (Kemal, 2019).

*Service Now Open-Source Ticketing System*

In an application, issues are normal that's why every support team role is very critical and in need. Incident ticket is categorized by two differentiations, the Auto and Critical Tickets. Auto tickets are like a notification. It is usually happening when the databases reach their limitations, the application is down, some processes are not running, and backup is failed. Critical tickets come from a user-raised issue and changes made by the support team. It usually tags is critical especially when the issue has affected the live production.

*Figure 1.* Service Now (SNOW) Open-Source Ticketing System

Figure 1 shows us an example of a ticketing system. It also includes whether the ticket is categorized as low, mid, high, urgent, or critical. Auto tickets are usually coming as low and mid tickets and it doesn't have any impact on the production or the user of the application but it affects the productivity of the support team. It serves as the notification alert for the support team to perform a health check for the application. Sometimes low tickets have been ignored because it is just an alert and nothing will happen. Mid tickets are usually the tickets seen by the site and discuss with the support team to prevent causing impact especially when the warehouse is on production. As long as the issue comes with no impact on production or it doesn't affect users', tickets can be tagged as low or mid depending on their description.

When the production experiences a small quantity of impact, for example, one (1) or two (2) users are unable to access the application, it comes with a high ticket. Some ticket especially if it is related to application access, databases, or caused by a change made by the support team it is subject to becoming critical if the support team doesn't take any actions to prevent or troubleshoot the situation as early as possible. When a user raises or escalates an issue and it has affected the live production, the ticket is tagged as critical. Aside from user raised, Changes made by the support team are sometimes the cause why there's a high or critical ticket. Changes are usually executed to maintain, upgrade, reboot, update, etc. the application. It helps also to refresh its settings and just like a human, to rest. When the issue comes from the changes made by the team and it has a large quantity of impact on the users, the ticket is tagged as critical. Although issues from the change are also a user raised because it is usually coming after the validation, the difference is change come from the support teams' end while on the other hand, its users' error or something happened to the application itself (e.g., power or network outages).

*Why do researchers use Service Now Open-Source Ticketing System?*

Service now is popular as a ticketing system and it has a lot of features. The researchers are currently working in an IT Industry that uses this tool as a service provider for their clients. Researchers are already familiar with how the service now works from the creations of different tickets, creating dashboards and task boards, and other service configurations such as SLA and KM/KB. Because of familiarity, researchers decided to use Service Now Open-Source Ticketing system to be the main source of ticket data and ticketing system tool.

## METHODOLOGY

The researcher used a data collection method for this specific study. The data comes from the service now (SNOW) open-source ticketing system and purposive sampling that has been analyzed to highly represent the data. This also shows the possible impact of every ticket's category. Data collection is a process of collecting and measuring data from different people but with the same variable category. In every research, data collection or gathering is common. It is usually the first step to have any information to analyze the problem and think for a possible solution.

The objective for data collection is to produce high-quality proof of data that can lead us to data analysis and to build conclusive and reliable answers to every question (Gilder & Schofield, 1978). Open source refers to something that everyone can alter and share. Simply it is designed for public use. This has been used for this research to give an idea for ticketing system in following the data privacy act (Philippine Law "Republic Act No. 10173") and data privacy law for every company that is using ticketing system and also for their security (Batara & Flora, 2017; Tse, 2014). Survey research is also being used by the researcher for data collection. Its purpose is to analyze the experience of support teams in handling tickets. This is also a technique to gather information from different people but with the same expertise. The goal of this survey is to acquire enough data samples that will represent the population of interest (Lichtman, 2011).

Appendix A shows the questions that have been asked to the survey. The researcher conducts purposive sampling from ABC company (For data privacy issue, unable to disclose the name of the company) and chooses fifty (50) different people who are working as a Technical/Application Support. This questionnaire focuses on their experiences in handling the ticket using a ticketing system, end to end, and how auto and critical tickets affect their work productivity. It also asked suggestions on how to deal with both tickets.

## RESULTS AND DISCUSSION

The main objective of the research is to identify the possible business impact of the auto and critical ticket both users raised and auto-generated and provide a solution.

Using this research, companies or businesses that use ticketing systems will have an idea of the productivity of their resources. This will also help the management to have better planning on how to resolve and identify properly the tickets on its priority and what ticket should they do first.

## *AUTO TICKET AND CRITICAL TICKET IN LEVEL 1 QUEUE*

| Number ▼ | Opened | Short description | Caller | Priority | State | Category | Assignment group |
|---|---|---|---|---|---|---|---|
| INC0010015 | 2020-12-07 08:31:42 | Process Not Running | Barbara Hindley | 3 - Moderate | New | Software | L1 Queue |
| INC0010014 | 2020-12-07 08:31:02 | Backup Failed | Vivian Brzostowski | 4 - Low | New | Hardware | L1 Queue |
| INC0010013 | 2020-12-07 08:30:31 | Database utilization exceeds 90% threshold | Vince Ettel | ● 1 - Critical | New | Database | L1 Queue |
| INC0010012 | 2020-12-07 08:29:31 | Database utilization exceeds 75% threshold | Avery Parbol | ● 2 - High | New | Database | L1 Queue |

*Figure 2.* Auto Tickets and Critical tickets in Level 1 Queue

Figure 2 shows us the auto tickets that the system generated. Auto tickets are the auto-generated tickets that the system produces to give an alert to the resource that the system will have or have a problem. With the help of auto tickets, resources can do proactive checking on the server or application. In the article of Blundo, Arco, De Santis, and Galdi (2002), they emphasize the importance of doing proactive checking. Although their article is related to password pro-active checking, simply doing a pro-active check helps everyone to prevent a possible impact on the server or application (Blundo et al., 2002).

The server or application generates an auto ticket especially when the error is much related to software. For example, one of the auto tickets in Fig. 3 is the Backup Failed. Backup Failed means the alternative or the second option failed to run again. A possible error is the tape for backup is not on the tape drive. As we identify the problem, auto tickets directly go to the level 1 queue. Level 1 resources tasks are to monitor the queue and update all tickets. They need to make sure that all tickets' next steps are for monitoring or not on their task. A sudden issue like power outage and network outage is also going to Level 1 queue. This kind of Critical Ticket is system generated and it impacts the system or application directly. When it happens, Level 1 resources are going to perform basic checkings, and once it is confirmed that the system or application is down, immediately raise the issue to Level 2 resources to perform further checkings and identify the root cause. With this, it comes the SLA or Service Level Agreement.

Service Level Agreement is the contract between the Service Providers and the Customers that identifies the time frame for the service providers on how long the ticket or issue must be resolved. Usually, it depends on the ticket's priority. This is between the Providers' and Customers' discussion on how long the Critical, High, Mid, and Low priority tickets are to be resolved or open. Once the ticket becomes out of SLA, service providers need to pay for the damages or they will take the penalties of not meeting their committed goals (Marilly et al., 2002). Service Level Agreement helps not only the customer but also the Service Provider. When the Service Provider's percentage for SLA is high, it means that they always meet the customers' expectations and resolve the critical ticket on or before its SLA time. This will attract more companies to sign a partnership and increase their profit. On the Customers' side, it guarantees them that once they have a critical issue it will be resolved by the given time and will help them analyze and create planning to lessen the possible business impact of the said critical issue. One of the tasks of the Level 1 resource is to monitor the L1 Queue. They need to make sure that all new auto-tickets are being checked immediately once it comes to the queue.

In the survey, about productivity in handling auto tickets, surveillant 1 said *"Auto generated tickets have details on it (description incident type, etc.), so it is easy to determine what kind of issue/error encountered and what should be the resolution needed to resolve it".* For him, it is easy to work on it cause the problem is already indicated on the ticket as this is auto-generated by the system or application. Surveillant 2 said, *"For auto-tickets, it depends on the priority but mostly it consumes quite a time and effort but if you're familiar with the issue you can pretty solve it quicker than the first time".* Because auto tickets came to the queue, he can resolve it quickly cause it happens every time and it doesn't affect his productivity cause he already knows how to resolve the issue. Also, he said, *"Just like stated earlier auto-tickets help to prevent a much higher issue. So we can have more time to work on user-raised tickets or high priority tickets".*

From the data gathered, Level 1 resources work productivity in dealing with auto tickets is depends on how he or she is familiar with the issue. As this issue frequently comes to the queue, Level 1 faces it almost every day and does the same fix. Aside from the frequent tickets, they are also monitoring new tickets coming from the queue. We can identify that once the critical ticket comes, all auto-ticket will be ignored as the critical ticket must resolve first in line with the service level agreement. Surveillants also identified that auto-tickets does not affect much real-time production but in a long run, it can. For example, in Figure 3 ticket 1, Process not running. From its problem not running, it means that some process of the system or application has been halted and it may affect the connection. It does not affect the real-time production on the spot but if that process will not run it may lead to some connection error or users end error. With the help of Auto-Tickets, the system will generate a ticket to the queue to have given an alert for level 1 to make a proactive check to prevent any possible critical tickets. Surveillant 2 also said how auto ticket affects the users *"Yes, it helps the support team to prevent a much higher issue".*

| | | Number ▼ | Opened | Short description | Caller | Priority | State | Category | Assignment group | Assigned to | Updated | Updated by |
|---|---|---|---|---|---|---|---|---|---|---|---|---|
| | ⓘ | INC0010015 | 2020-12-07 08:31:42 | Process Not Running | Barbara Hindley | 3 - Moderate | New | Software | L1 Queue | (empty) | 2020-12-07 08:32:08 | admin |
| | ⓘ | INC0010014 | 2020-12-07 08:31:02 | Backup Failed | Vivian Brzostowski | 4 - Low | New | Hardware | L1 Queue | (empty) | 2020-12-07 08:31:36 | admin |
| | ⓘ | INC0010013 | 2020-12-07 08:30:31 | Database utilization exceeds 90% threshold | Vince Ettel | 1 - Critical | New | Database | L1 Queue | (empty) | 2020-12-07 08:30:56 | admin |
| | ⓘ | INC0010012 | 2020-12-07 08:29:31 | Database utilization exceeds 75% threshold | Avery Parbol | 2 - High | New | Database | L1 Queue | (empty) | 2020-12-07 08:30:19 | admin |
| | ⓘ | INC0010004 | 2020-12-07 08:24:27 | Network Outage | Abraham Lincoln | 1 - Critical | New | Network | L1 Queue | (empty) | 2020-12-07 08:35:58 | admin |
| | ⓘ | INC0010003 | 2020-12-07 08:21:45 | Power Outage | Abel Tuter | 1 - Critical | New | Inquiry / Help | L1 Queue | (empty) | 2020-12-07 08:36:10 | admin |
| | ⓘ | INC0007002 | 2018-10-16 22:47:51 | Need access to the common drive. | David Miller | 4 - Low | New | Inquiry / Help | L1 Queue | (empty) | 2020-12-07 08:36:38 | admin |
| | ⓘ | INC0001990 | 2020-06-07 09:02:25 | Unable to access the personal details section in payroll portal | Problem CoordinatorATF | 5 - Planning | On Hold | Inquiry / Help | L1 Queue | (empty) | 2020-12-07 08:37:10 | admin |
| | ⓘ | INC0000059 | 2016-08-10 09:14:29 | Unable to access team file share | Rick Berzle | 3 - Moderate | New | Inquiry / Help | L1 Queue | (empty) | 2020-12-07 08:37:20 | admin |
| | ⓘ | INC0000054 | 2015-11-02 12:49:08 | SAP Materials Management is slow or there is an outage | Christen Mitchell | 1 - Critical | On Hold | Software | L1 Queue | (empty) | 2020-12-07 08:38:14 | admin |
| | ⓘ | INC0000049 | 2020-09-03 14:56:37 | Network storage unavailable | Beth Anglin | 2 - High | In Progress | Network | L1 Queue | (empty) | 2020-12-07 08:39:54 | admin |
| | ⓘ | INC0000041 | 2020-06-26 17:44:53 | My desk phone does not work | Bow Ruggeri | 3 - Moderate | In Progress | Hardware | L1 Queue | (empty) | 2020-12-07 08:40:34 | admin |
| | ⓘ | INC0000040 | 2020-06-26 17:42:45 | JavaScript error on hiring page of corporate website | Bud Richman | 3 - Moderate | On Hold | Inquiry / Help | L1 Queue | (empty) | 2020-12-07 08:40:49 | admin |
| | ⓘ | INC0000039 | 2020-06-26 17:41:01 | Trouble getting to Oregon mail server | Bud Richman | 5 - Planning | New | Network | L1 Queue | (empty) | 2020-12-07 08:41:03 | admin |
| | ⓘ | INC0000037 | 2020-06-26 17:34:56 | Request for a new service | Sam Sorokin | 3 - Moderate | In Progress | Inquiry / Help | L1 Queue | (empty) | 2020-12-07 08:41:13 | admin |
| | ⓘ | INC0000029 | 2020-06-24 17:00:44 | I can't get my weather report | Charlie Whitherspoon | 5 - Planning | In Progress | Inquiry / Help | L1 Queue | (empty) | 2020-12-07 08:41:44 | admin |
| | ⓘ | INC0000027 | 2020-06-22 16:55:55 | Please remove the latest hotfix from my PC | Fred Luddy | 2 - High | In Progress | Software | L1 Queue | (empty) | 2020-12-07 08:41:54 | admin |
| | ⓘ | INC0000025 | 2020-06-03 16:53:46 | Need to add more memory to laptop | Don Goodliffe | 2 - High | In Progress | Hardware | L1 Queue | (empty) | 2020-12-07 08:46:16 | admin |

*Figure 3.* Level 1 Queue

Figure 3 shows us the Level 1 Queue where the auto-generated tickets of the system or application are sent. This is also the queue where Level 1 resources are monitored.

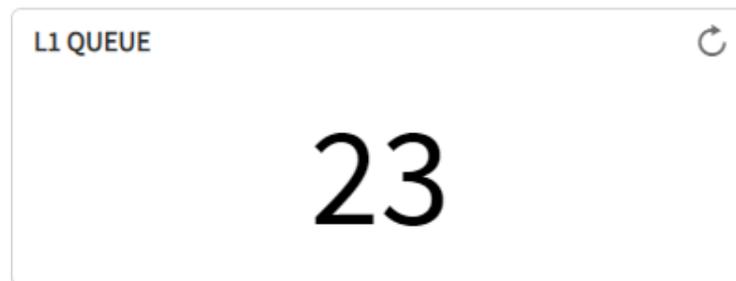

*Figure 4.* Level 1 Queue view in Dashboard

Figure 5 shows us the Level 1 Queue view in Dashboard. It helps to give a quick knowledge for Level 1 resource on how many tickets are still there on their queue and also for other teams who want to check the tickets on the said queue.

### USER RAISED AND CRITICAL TICKET LEVEL 2 QUEUE

Figure 5 shows the user-raised and critical ticket. Unlike in Level 1, Auto-generated tickets go to Level 1 Queue, User raised comes to this queue. In this queue, Level 2 resources are entitled to help direct all tickets made by the user. INC0010011 User zz1111 is unable to log in. As this issue is by a user, the Level 2 resource task is to resolve this. One of the criteria or requirements of the Level 2 is at least he or she knows how to identify and do further checkings about the issue. In the example, user zz1111 is unable to log in.

Level 2 must identify why it is unable to log in. He or she can check if the user id is suspended or it has a grace login. Level 2 resource task is to make sure that all user-raised tickets will be resolved.

| Number ▼ | Opened | Short description | Caller | Priority | State | Category | Assignment group |
|---|---|---|---|---|---|---|---|
| INC0010011 | 2020-12-07 08:28:49 | User zz1111 unable to login | Annie Approver | 3 - Moderate | New | Software | L2 Queue |
| INC0010010 | 2020-12-07 08:28:20 | All user unable to login | Andrew Och | ● 1 - Critical | New | Software | L2 Queue |
| INC0010009 | 2020-12-07 08:27:40 | Printer Zzz unable to print | Alfonso Griglen | 3 - Moderate | In Progress | Software | L2 Queue |
| INC0010008 | 2020-12-07 08:27:08 | All Printers Unable to Print | Angelo Ferentz | ● 1 - Critical | New | Software | L2 Queue |
| INC0010007 | 2020-12-07 08:26:43 | IDOCs unable to send Both Flow | Alene Rabeck | ● 1 - Critical | New | Inquiry / Help | L2 Queue |
| INC0010006 | 2020-12-07 08:26:00 | IDOCs unable to send Outbound | Aileen Mottern | ● 2 - High | New | Software | L2 Queue |
| INC0010005 | 2020-12-07 08:25:24 | IDOCs unable to send Inbound | Adela Cervantsz | ● 2 - High | New | Software | L2 Queue |

*Figure 5.* User Raised and Critical Ticket in Level 2 Queue

In having a critical ticket, L2 must lead the entire process. He or she is the one that must do the technical troubleshooting as this is direct to the system or application. Aside from that, L2 must also know that the critical ticket must be resolved in a line with the service level agreement. The same with the L1, L2 queue will be left behind once there's a critical ticket as this is the top priority. One of the incidents issues in fig. 6 is the IDOC's not flowing inbound. It is has a priority of high cause it affects the direction of the system and it has a possible business impact. In this case, L2 must take over cause it is a user-raised ticket. Another incident in fig. 6 is IDOC's not flowing both ways it means that inbound and outbound IDOC's are not flowing or sending from Application to SAP and SAP to Application. It has a critical priority cause it affects all users and it has a possible business impact. Critical ticket is usually identified if it is affecting all users and the system like the power outage and network outage. The difference between those critical tickets are power outages and network outages are auto-generated issues by the system or application itself while the IDOC's not flowing both ways is user raised but still if it is categorized as critical priority L2 must take over and do the technical troubleshooting with the help of the L1 for documentations.

In the survey, surveillant 3 said regarding how critical ticket affects his productivity *"Not much, dealing with tickets is part of or productivity whether it's critical or not".* It is pretty easier to resolve the incident as long as you know who and what team you will contact. Surveillant 4 also said, *"I can only think that quantity of tickets affects the work productivity as it's our job to deal with it".* As L2 is the one who troubleshoots all the user-raised tickets, it means that if they are having a critical incident they need to put it as the priority to resolve. The same thing we said a while ago. Also because of that more or less they will be overwhelmed by other tickets. It is also seconded by surveillant 5, as for him *"Since managing critical issue warrant your utmost attention, I'm not able to multitask or perform other deliverables".*

From the data gathered, Level 2 resources work productivity in dealing user raised tickets is depends on how he or she is fast to analyze what is the problem. Aside from that Level 2 monitors the L2 queue if there's a new user-raised ticket comes. Unlike Level 1, tickets are the same every day caused it is an auto ticket generated by the system and is a default error. In Level 2 it is not the same every day cause it always depends on what the user encountered a problem. Figure 6 shows us the Level 2 Queue where the user-raised ticket goes. This is also the queue where Level 2 resources are monitored.

*Figure 6.* Level 2 Queue

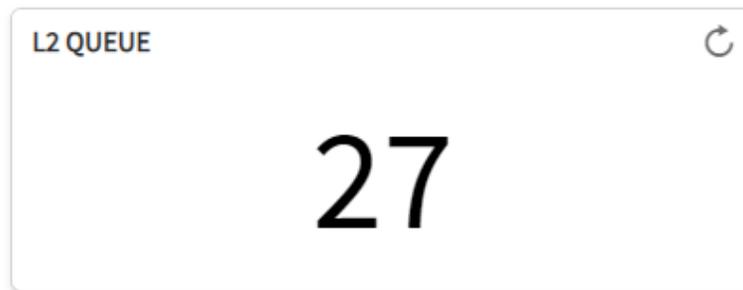

*Figure 7.* Level 2 Queue view in Dashboard

Figure 7, shows us the Level 2 Queue view in Dashboard. It helps to give a quick knowledge or background for Level 2 resource on how many tickets are still on their queue and for other teams who wants to check the tickets on the said queue.

## LEVEL 3 QUEUE

Level 3 Queue or L3 Team, is one of the highest technical levels in support teams. They are usually helpful once the critical ticket is already out of service level agreement. Aside from that Level 3's tasks are more on Problem Management the next step after resolving the ticket.

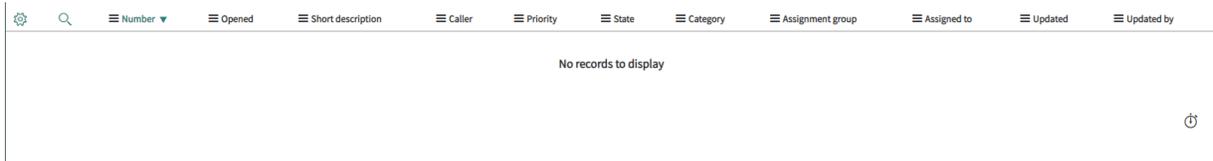

*Figure 8.* Level 3 Queue

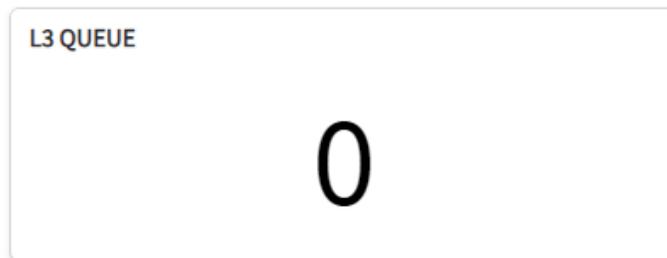

*Figure 9.* Level 3 Queue view in Dashboard

Figure 8 shows us the Level 3 Queue but as per the data, there is no ongoing ticket with it the same thing with Figure 9. Figure 9 is the view in Dashboard. It helps to give an easy view of how many tickets are still ongoing with the Level 3 queue.

Figure 10 shows the Dashboard of Auto and Critical Tickets along with its priority from Critical to Low state priority. Aside from the bar graph, it also shows the exact count of tickets depending on its category of assignment group. We can see the Open Incident, Critical Open Incident, L1 Queue, L2 Queue, L3 Queue, Closed Ticket, and Resolved Ticket. A dashboard is a tool that allows the alignment between two or more business processes. It provides accurate information to allow the management or analyst to analyze and create a decision and compare the result with the planned goal. Having too much data and analyze manually by a person can lead to misinformation and inaccurate result, dashboard becomes very effective as it gathers all the data and shows it simply using a bar graph and numeric. Although, there are still many user interfaces for the dashboard depending on the one who creates it. Aside from that, dashboards answer easily the different questions of different audiences (Murnawan, Samihardjo, & Nugraha, 2020).

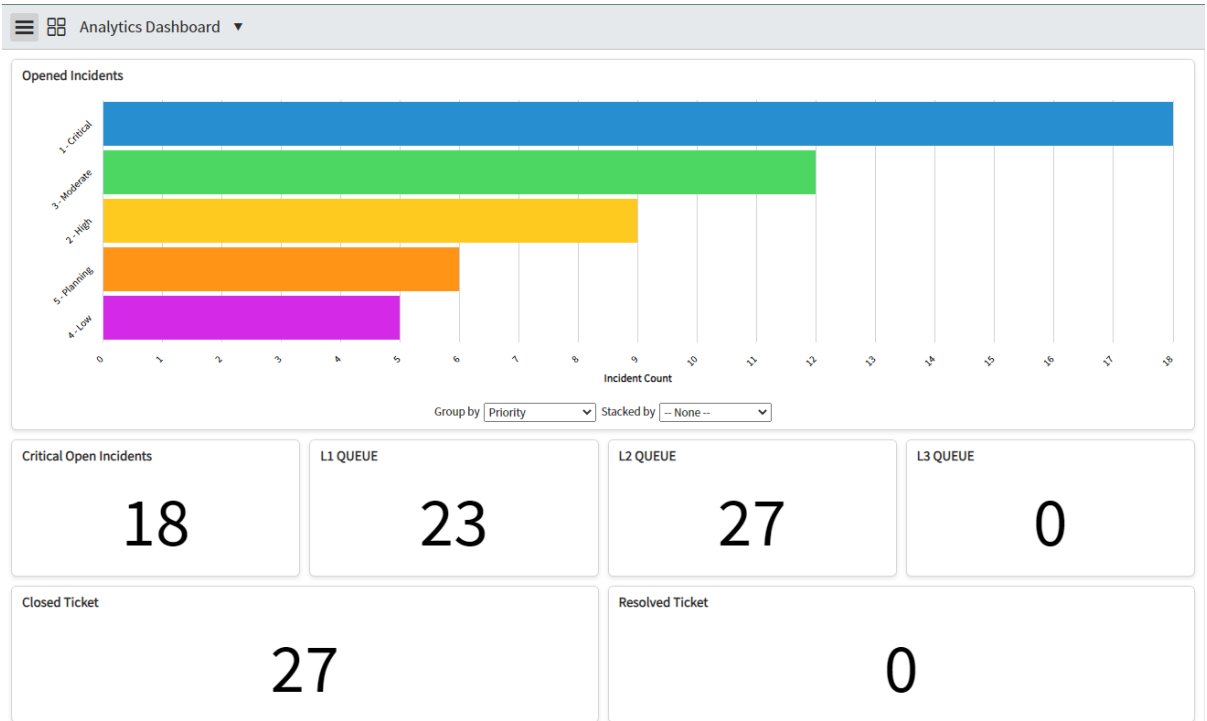

*Figure 10.* Analytics Dashboard

      Different audiences like L1, L2, L3, or even Managers who want to take a look at the current count and priority of the tickets can use the dashboard to easily have an idea whose queue and resources have a lot of work to do. With this, management can create a strategy and analysis on how auto and critical tickets affect the productivity of the resources. Critical tickets are viewed in Figure 10 which gives the management an overview of what is going on in the real-time production. It can help them to identify the business impact of the said critical tickets and make a plan on how to prevent possible critical issues. It also helps the resources, especially with their productivity.

      In Figure 10, Analytics Dashboard, we can see that there are 18 critical tickets open. This means that these 18 tickets are needed to be resolved as soon as possible in line with the service level agreement between the provider and consumers. As we discussed in 3.1 Auto and Critical Ticket in Level 1 Queue, if the provider didn't meet the SLA, they need to pay or they are charged with the consequence in the contract and it is a loss within the company. Because these 18 critical tickets open, L1 and L2 Queue will be on hold and it is possible to have more tickets in a short period while the resources are troubleshooting and resolving the critical opened tickets. One of the impacts of having critical tickets is the resources are not able to do multitask. This was also from surveillant 5. Once the critical tickets have been resolved, the next thing is the tickets on the queue. With this, we can analyze that the resource has a lot of work to do and can affect their productivity. Also, it affects real-time production because user-raised tickets and auto-tickets will be on hold. Remember that, user-raised is directly from the customers and auto-tickets are auto-generated issues from the system or application that will help to prevent possible

critical or long-term business impact. Ignoring the user-raised will affect the relationship between the provider and the customer and it is another loss to the company while the auto tickets are possible long-term business impact.

## *Survey (Purposive Sampling)*

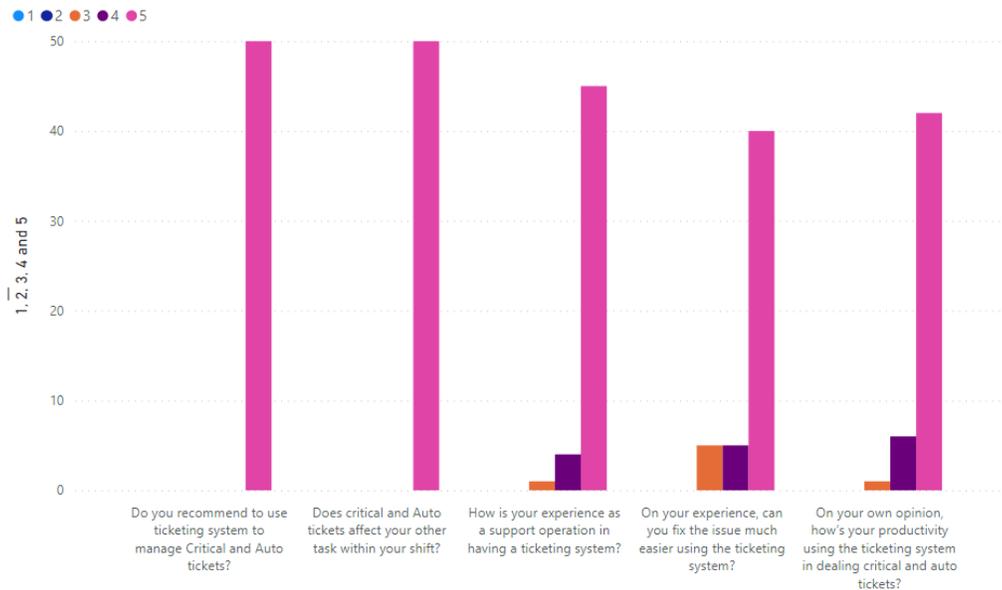

*Figure 11.* Clustered Column Chart Result of Survey

Figure 11 shows us the result of the survey. In general, almost all the surveillants choose 5 as this is the highest. It means that Critical and Auto Tickets affect their productivity in an everyday transaction. In question 1 and 2, the results are, having a ticketing system increase their productivity as they can fix the escalated issue and request. They have a reference on how to fix it and monitor the status of the ticket. In question 3, the result is, critical and auto ticket affects them as they can't maximize their time. It means that handling critical and auto tickets is time-consuming especially without any reference. It also affects the progress of the team especially onboarding new employees and at the same time troubleshooting critical and auto tickets. Question 4, still, majority of the votes that ticketing system helps to maximize their productivity rather than having none. And lastly, in question 5, all of the surveillants recommends having a ticketing system to monitor and document critical and auto ticket.

Researchers also asked the surveillant for their suggestions on how to deal with critical and auto ticket, and most of them suggest that having Knowledge Management or Knowledge-Based maximize their productivity as they will have a reference on how to perform checking, fix the issue, and onboard new employees.

# CONCLUSIONS AND RECOMMENDATIONS

Research for Impact of Auto and Critical ticket is simply to have an analysis for both Management and Application Support Workers to have an idea and create a strategy on how to manage the Auto and Critical tickets that they may encounter. Service Now Open Source ticketing system and conducting survey helps the researchers to visualize and create an analysis on how these two-incident tickets are different from each other and affect the business in real-time and in a long run. It is concluded that Auto and Critical tickets can affect the resource work productivity, provider and consumers relationship, and business value. As critical tickets are the top priority, it affects the availability of the resources to multi-task. Because of that, queues will have pending tickets that need to be resolved. It affects the productivity of the resource because they will be tired of resolving the critical incident and after that still too many tickets are pending.

Critical and other high-priority tickets need to be resolved per the service level agreement between provider and consumer. If the provider doesn't meet the service level agreement it affects its relationship with the customer and it will lead to business value impact or business loss. It is concluded that everything will fall on how the management creates a plan for production flow and on how the resources will manage auto and critical tickets. In line with this, it is recommended to have Knowledge Management/Knowledge-Based that all Levels will use to troubleshoot issues. Knowledge Management/Knowledge-Based or KM/KB is a compilation of documents. If the application support workers have a tool of KM/KB it is easy to troubleshoot all ongoing and upcoming auto or critical tickets. Not only that, it is not hard for the senior resource to guide and teach the new resources especially if the new resource doesn't have any troubleshooting or support background. Aside from that, having KM/KB will also help the Level 1 and Level 2 resources to resolve all auto and user raised tickets as quickly as possible to avoid having too many tickets on their queues and focus if there's a critical ticket as they need to resolve it per SLA. Aside from KM/KB, it is also recommended to have documentation of all critical incidents. This will help all resources including the management that there's an ongoing critical issue so everyone will be alert of its possible business impact.

This is another way to have a root cause analysis to prevent such critical tickets especially if the critical issue is technology-related. Root cause analysis is another recommendation as this can help the management to easily audit the business impact. This research is recommended to present to different IT companies that use ticket systems in their data center. This is helpful especially for both live production and a management-related team that supports manufacturing companies and real-time application consumer use. Lastly, for further study, researchers will survey other IT companies who are using or not a Ticketing System.

# Appendix A
Survey Questionnaire

This questionnaire will be used as reference and information for data analysis. This will also include in the article entitled: *"Impact of Critical and Auto Tickets: Analysis for Management and Worker's Productivity in using Ticketing System"*.

Do you agree to be used your answers for an article?

If yes, please continue to answer the following questions and fill the code name for data privacy.

Code Name:

5 – Highest; 1 – Lowest

| | 5 | 4 | 3 | 2 | 1 |
|---|---|---|---|---|---|
| How is your experience as a support operation in having a ticketing system? | | | | | |
| On your experience, can you fix the issue much easier using the ticketing system? | | | | | |
| Does critical and Auto tickets affect your other task within your shift? | | | | | |
| On your own opinion, how's your productivity using the ticketing system in dealing critical and auto tickets? | | | | | |
| Do you recommend to use ticketing system to manage Critical and Auto tickets? | | | | | |

Please give some suggestions on how to deal or troubleshoot critical tickets by having a good work productivity or less time consuming.